\documentclass[twocolumn,prc,showpacs,preprintnumbers,amsmath,amssymb,superscriptaddress]{revtex4-2}

\usepackage{graphicx,epsfig}% Include figure files
\usepackage{dcolumn}% Align table columns on decimal point
\usepackage{bm}% bold math
\usepackage{tabularx}
\usepackage{ulem}
\newcolumntype{M}{>{\centering\arraybackslash}m{1.85cm}}
\usepackage[export]{adjustbox}
\usepackage{float}
\usepackage{color}   %May be necessary if you want to color links
\usepackage{hyperref}
\hypersetup{
	colorlinks=true, %set true if you want colored links
	linktoc=all,     %set to all if you want both sections and subsections linked
	linkcolor=blue,  %choose some color if you want links to stand out
}
\usepackage{placeins}
\usepackage{booktabs}
\usepackage{blindtext}
\makeatletter
\newcommand{\colorcaption}[2][]{%
	\begingroup%
	\renewcommand{\@caption@fignum@sep}{ (Color online). }%
	\caption[#1]{#2}%
	\endgroup%
}
\makeatother
\begin{document}
\title{A unified description of the structure and electromagnetic breakup of $^{\bf 11}$Be}
\author{M. Dan}
\email{mdan@ph.iitr.ac.in}
\affiliation{Department of Physics, Indian Institute of Technology Roorkee, Roorkee 247 667, India}
\author{R. Chatterjee}
\email{rchatterjee@ph.iitr.ac.in}
\affiliation{Department of Physics, Indian Institute of Technology Roorkee, Roorkee 247 667, India}
\author{M. Kimura}
\email{masaaki@nucl.sci.hokudai.ac.jp}
\affiliation{Department of Physics, Hokkaido University, 060-0810 Sapporo, Japan}
\affiliation{Nuclear Reaction Data Centre, Faculty of Science, Hokkaido University, 060-0810 Sapporo, Japan}
\affiliation{Research Center for Nuclear Physics (RCNP), Osaka University, Ibaraki 567-0047,    Japan}

%\author{C}
%\email{C}

%\author{V. Choudhary}
%\email{Put your official email address}
%\affiliation{Department of Physics, Indian Institute of Technology Roorkee, Roo%rkee 247 667, India}
%\author{W. Horiuchi}
%\email{whoriuchi@nucl.sci.hokudai.ac.jp}
%\affiliation{Department of Physics, Hokkaido University, 060-0810 Sapporo, Japa%n}
%\author{M. Kimura}
%\affiliation{Department of Physics, Hokkaido University, 060-0810 Sapporo, Japa%n}
%\affiliation{Nuclear Reaction Data Centre, Faculty of Science, Hokkaido Univers%ity, 060-0810 Sapporo, Japan}
%\affiliation{Research Center for Nuclear Physics (RCNP), Osaka University, Ibara%ki 567-0047, Japan}         
%\author{R. Chatterjee}   
%\affiliation{Department of Physics, Indian Institute of Technology Roorkee, Roo%rkee 247 667, India}

% \date{\hfill \today}
	%%%%%%%%%%%%%%%%%%%%%%%%%%%%%%%%%%%%%%%%%
	
	\begin{abstract}
        We study both the static properties of $^{11}$Be and its reaction dynamics during electromagnetic
 breakup under a unified framework. A many-body approach - the antisymmetrized molecular dynamics
 (AMD) is used to describe the structure of the neutron-halo nucleus, $^{11}$Be. The same AMD wave
 function is then adapted as an input to the fully quantum theory of Coulomb breakup under the
 aegis of the finite range distorted wave Born approximation theory. The calculated observables are
 also compared with those obtained with a phenomenological Woods-Saxon potential model wave
 function. The experimental core-valence neutron relative energy spectrum and dipole response along
 with other observables are well described by our calculations. 
	\end{abstract}

\maketitle
\section{introduction}

Since the discovery of halo nuclei \cite{Tanihata halo}, several observations in their study  have shown unconventional results, which were contrary to traditional nuclear structure estimations. For example, unlike the case of stable nuclei where the matter radius generally follow the charge radius, the matter radius of $^{11}$Be was found to be larger than its charge radius.
The full width at half-maximum (FWHM) of the parallel momentum distribution (PMD) of stable nuclei ($\approx$ 140 MeV/c ) is much higher compared to that from a halo nucleus ($\approx$ 40 MeV/c). For $^{10,11,12}$Be breaking up on a heavy target (Au), the FWHM of the PMD of the charged fragment are 191.13 MeV/c, 43.23 MeV/c and 88.93 MeV/c, respectively \cite{Shubh}. 
%If we compare the FWHM for Be isotopes breaking up on heavy target then the values are 191.13 MeV/c, 43.23 MeV/c, 88.93 MeV/c for $^{10,11,12}$Be \cite{Shubh}, respectively. % Similarly, there exists a substantial amount of electric dipole strength at low excitation energies ($\approx$ 1 MeV) for halo nuclei, whereas the dipole strength for stable nuclei, known as giant dipole resonanace, are at an excitation energy of 15 - 20 MeV.

It is generally considered that a large  neutron to proton ratio results in a sharp decrease of the
one-neutron separation energy, and consequently an extension in the neutron wave function far outside
the nuclear mean field is observed \cite{Aumann}. This extension directly affects the static
properties of the system. The root mean square matter radius of $^{10}$Be and $^{11}$Be are 2.30
$\pm$ 0.02 fm and 2.73 $\pm$ 0.05 fm, respectively \cite{Tanihata}. In a neutron halo nucleus, one can get
information on the interaction between the clusters of a dicluster nucleus from the change of
nuclear charge distribution. These changes may occur due to the relative motion between the core and the 
centre of mass, and due to the core polarization resulting from the core-valence neutron interaction
\cite{Charge radii}. There are several studies \cite{Wang, Suhelahmed, Krieger, NCSM} that report on charge and matter radius of
$^{11}$Be. However more analysis are required for a consistent picture. The comparison between charge and
matter radii is significant for the nuclei with different distribution of neutron and proton halo. 

Tanihata {\it et. al.} \cite{Tanihata halo} have shown that there is a notable increase in interaction cross section for drip line nuclei compared to the neighbouring isotopes of light elements. There is an increase in one neutron removal cross section ($\sigma_{n}$) for $^{11}$Be as compared to $^{10}$Be, while breaking up on a Pb target. The average one neutron removal cross section for $^{10}$Be (beam energy ranging from 37 - 70 MeV/u) is 0.126 $\pm$ 0.011 b, while that for $^{11}$Be (beam energy ranging from 17 - 66 MeV/u), is 2.16 $\pm$ 0.17 b \cite{Warner}. It is evident that the average one neutron removal cross section of $^{11}$Be is an order of magnitude higher than that of $^{10}$Be. The unusually large reaction cross section of a halo nuclei, compared to its isobars, is a consequence of  the matter radius significantly deviating from the usual $A^{1/3}$ dependence  expected for stable nuclei \cite{Capel2018}.  

The analysis of an external nuclear or electromagnetic field response by a nucleus is one of the key elements to understand the characteristics of a many-body nuclear system \cite{Lei}. At present, there are several discussions about the way giant dipole resonance strength evolves from stable to weakly bound exotic nuclei in extreme neutron to proton ratios. In general, the presence of collective soft-dipole resonance is expected to occur in heavier neutron-rich structures at excitation energies lower than the giant dipole resonance \cite{softdipole1,softdipole2}. Such a mode may arise when loosely bound valence neutrons vibrate against the residual core. In the literature, it is often referred to as pygmy resonance. In electromagnetic dissociation experiments (\textit{e.g.} \cite{NakamuraPLB, NakamuraPRL}), a prominent low-lying dipole strength was observed in light halo nucleus. Their presence is justified by two arguments: first is due to the coherent vibration of two halo neutrons against the charge core ($\textit{e.g.}$ $^{6}$He \cite{6He} and $^{11}$Li \cite{11li}) and second is due to the non-resonant breakup of one neutron halo nucleus ($\textit{e.g.}$ $^{11}$Be \cite{NakamuraPLB} and $^{19}$C \cite{NakamuraPRL}) into the continuum \cite{Lei, BertulaniNPA, BertulaniPRC}. 
It is also reported that astrophysical aspects such as abundance pattern in the r-process
nucleosynthesis could also be related to the presence of low-lying dipole strength present in
neutron-rich nuclei \cite{Goriely,BertulaniEPJA}.  

In this article, we aim to combine nuclear structure and reaction models to discuss both the static
and dynamical properties of a neutron-halo nucleus, $^{11}{\rm Be}$. For this purpose, we use
the antisymmetrized molecular dynamics (AMD) \cite{kanada2003,kanada2012,kimura2016} to calculate
the static properties such as 
one neutron separation energy, charge and matter radii. The AMD wave function is also used 
as an input to the fully quantum mechanical Coulomb breakup theory of finite range distorted wave
Born approximation (FRDWBA) to calculate several reaction observables in the breakup of $^{11}$Be
on a heavy target ($^{208}$Pb) such as triple differential cross section, neutron energy
distribution, parallel momentum distribution, relative energy spectrum, and dipole response of
$^{11}$Be. The results are also compared with the available experimental data, and also with those
obtained from a phenomenological wave function derived using a Woods-Saxon (WS) potential whose
depth is adjusted to fit the one neutron separation energy of $^{11}$Be. 

In the following section, a brief description of the FRDWBA theory and details of the AMD framework
are presented. The results and analysis from our calculations have been discussed in section 3,
wherein we present the static properties of $^{11}$Be followed by calculations of various reaction
observables in during its electromagnetic breakup on a heavy target.  The conclusions of our work
appear in section 4.

\section{Formulation}
\label{sec:1}
\subsection{Framework of FRDWBA} 
\label{amd.sec1}
If we assume a projectile \textit{`a'} ($^{11}$Be), consisting of substructures \textit{`b'} ($^{10}$Be) and `\textit{c}' (neutron) to breakup in the pure Coulomb field of a heavy target \textit{`t'} ($^{208}$Pb). Then, the triple differential cross section for the process \textit{a} + \textit{t} $\longrightarrow$ \textit{b} + \textit{c} + \textit{t} can be written as, 
\begin{eqnarray}
\dfrac{d^3\sigma}{dE_{b}d\Omega_{b}d\Omega_{c}} = \dfrac{2\pi}{\hbar v_{at}}\rho{(E_{b},\Omega_{b},\Omega_{c})}\sum_{l,m}|\beta_{lm}|^{2},
\label{a2.1}
\end{eqnarray}
 
where, $v_{at}$ is the relative velocity between the \textit{a}-\textit{t} system in the entrance channel, $\rho{(E_{b},\Omega_{b},\Omega_{c})}$ is the three body final state phase space factor \cite{Fuchs}. The reduced transition amplitude in the post form FRDWBA, $\beta_{lm}$, for the breakup process is given by \cite{RCPPNP},
\begin{eqnarray}
\beta_{lm}(\textbf{q}_{b},\textbf{q}_{c};\textbf{q}_{a}) = \left\langle \zeta_{b}^{(-)}(\textbf{q}_{b},\textbf{r})
\zeta_{c}^{(-)}(\textbf{q}_{c},\textbf{r}_{c})\right\vert V_{bc}(\textbf{r}_{1}) \nonumber\\
 \times \left\vert
\phi_{a}^{lm}(\textbf{r}_{1})\zeta_{a}^{(+)}(\textbf{q}_{a},\textbf{r}_{i})\right\rangle.
\label{a2.2}
\end{eqnarray}

\begin{figure}[h]
\centering
\includegraphics[width=\linewidth,clip]{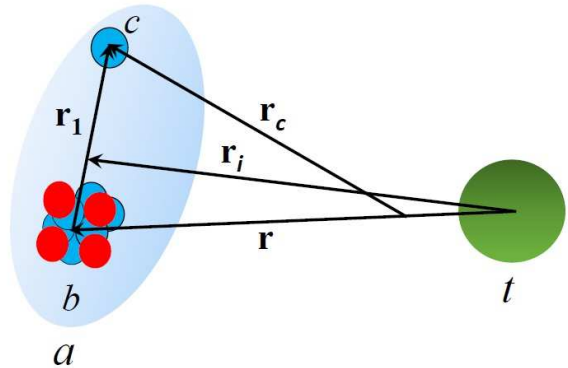}
\caption{The three body Jacobi coordinate system. The corresponding position vectors are denoted by \textbf{r}'s. }
\label{structure}
\end{figure}

$\zeta_{i}$'s (\textit{i = a, b, c}) are the pure Coulomb distorted waves of the appropriate particles with respect to the target and $\textbf{q}_{i}$s are the associated Jacobi wave vectors. The position vectors are shown in Fig. 1 with \textbf{r} = \textbf{r}$_i$ - $\alpha$\textbf{r}$_1$ and 
\textbf{r}$_c$ = $\gamma$\textbf{r}$_1$ + $\delta$\textbf{r}$_i$, where $\alpha$, $\gamma$ and $\delta$ are the mass factors:
$\alpha$ = m$_c$/(m$_c$ + m$_b$);  $\gamma$ =  m$_t$/(m$_b$ + m$_t$);  $\gamma$ = (1 - $\alpha\delta$ ) and m$_{i}$s (\textit{i = b, c, t}) are the masses of the appropriate particles.

$\phi_{a}^{lm}$({\textbf{r}$_1$}) is the ground state wave function of the projectile (\textit{a}), which is an eigenfunction of the two-body bound state potential $V_{bc}$(\textbf{r}$_1$),  with $l$ and $m$ as the orbital angular momentum  and its projection, respectively. Other reaction observables such as relative energy spectrum, neutron energy distribution, and parallel momentum distribution can be calculated by suitably integrating Eq. (1). The dipole strength distribution ${dB(E1)}/{dE}$ can be obtained \cite{ManjuEPJ} from the relative energy spectra ${d\sigma}/{dE_{rel}}$ using   
\begin{eqnarray}
\dfrac{d\sigma}{dE_{rel}}\ = \dfrac{16\pi^3}{9\hbar c}{n_{E1}}\dfrac{dB(E1)}{dE}\ ,
\end{eqnarray}
where,  $n_{E1}$ is the virtual photon number for electric dipole transition. 
For more details of the theory one is referred to \cite{RCPPNP}.

The main input to this theory is the wave function $\phi_{a}^{lm}$(\textbf{r}$_1$) = $u_l$({r}$_1$)
${Y^{lm}}$({\textbf{\^r}$_1$}), or more specifically the bound state radial wave function
$u_l$({r}$_1$). In this work, we test two different approaches to calculate this primary structure
input to the theory. An ordinary option is to calculate $u_l$({r}$_1$) from the core-valence neutron
interaction with the WS form whose depth is adjusted to reproduce the one-neutron separation energy with
fixed radius and diffuseness parameters. An alternative and more sophisticated approach would be to
use $u_l(\rm r_1)$ derived from a microscopic many-body wave function of the AMD, which is discussed
in details in the next section.        
    
\subsection{Model for the structure calculation} \label{amd.sec2}
The framework of AMD and the method to calculate valence neutron wave function (overlap amplitude)
are briefly explained. For more details, readers are directed to
Refs.~\cite{kanada2003,kanada2012,kimura2016}.  
\subsubsection{Framework of AMD} 
The Hamiltonian used in this study is given as,
\begin{eqnarray}
 H = \sum_{i=1}^A t(i) + \sum_{i<j}^A v_n(ij) + \sum_{i<j}^Z v_C(ij)  - t_{cm} ,
\end{eqnarray}
where, the Gogny D1S interaction~\cite{berger1991} is used as an effective nucleon-nucleon
interaction $v_n$. Following the prescription made in Ref.~\cite{homma2015},
we have weakened the strength of the spin-orbit interaction by 5\% from the original one to
reproduce the observed  splitting of the $1/2^\pm$ states of  $^{11}{\rm Be}$.  The Coulomb
interaction, $v_C$ is approximated by a sum of seven  Gaussians, and the center-of-mass kinetic
energy, $t_{cm}$ is exactly removed.  

The intrinsic wave function, $\Phi_{int}$  is represented by a Slater determinant of single-particle
wave packets. It is projected to the eigenstate of parity before the variation 
(parity projection before variation).
\begin{eqnarray}
 \Phi_{int}&=&{\mathcal A} \{\varphi_1\varphi_2\cdots\varphi_A \},\\
 \Phi^\pi_{int} &=& P^{\pi}\Phi_{int}=\dfrac{1+\pi \hat{P}_x}{2}\Phi_{int}, \quad (\pi=\pm),
 \label{eq:intwf}  
\end{eqnarray}
where, $P^\pi$ and $P_x$ denote the parity projector and parity operator, respectively. The
single-particle wave packet, $\varphi_i$ has the deformed Gaussian form~\cite{kimura2004},
\begin{eqnarray}
 \varphi_i({\bf r}) =&& \prod_{\sigma=x,y,z}
 \left(2\nu_\sigma/\pi\right)^{1/4}
 \exp\left\{-\nu_\sigma(r_\sigma -Z_{i\sigma})^2\right\}\chi_i\eta_i,\nonumber\\
 &&(i = 1,2,...A). 
 \label{eq:singlewf} 
\end{eqnarray}
$\chi_i$ is the nucleon spinor and $\eta_i$ is the isospin fixed to either of proton or neutron.  
The parameters of the wave function (${\bm Z}_i$, $\bm \nu$ and $\chi_i$)  are determined by the
energy variation which minimizes the expectation value of the Hamiltonian, 
\begin{eqnarray}
 H&=\dfrac{\langle \Phi^\pi|H|\Phi^\pi\rangle}{\langle
  \Phi^\pi|\Phi^\pi\rangle} +  v_\beta(\langle\beta\rangle-\beta)^2.
\end{eqnarray}
Note that the potential $v_\beta(\langle\beta\rangle-\beta)^2$ imposes the constraint on
the quadrupole deformation parameter $<{\beta}>$ defined in Ref. \cite{kimura2012}.
The magnitude of $v_\beta$ is chosen large enough so that $<{\beta}>$ equals to $\beta$.
By the energy variation, we obtain the optimized wave function for each given value of $\beta$,
which are denoted by $\Phi^\pi_{int}(\beta)$. No constraint was imposed on the other quadrupole
deformation parameter ${\gamma}$, and hence, it always has the optimal value.

\subsubsection{The generator coordinate method and AMD plus resonating group method} 
To describe the ground and excited states, we perform the angular momentum projection and
the generator coordinate method. The optimized wave functions $\Phi^{\pi}_{int}(\beta)$ are
projected to the eigenstates of the total angular momentum,
\begin{eqnarray}
 \Phi^{J\pi}_{MK}(\beta) &=& P^{J}_{MK}\Phi^{\pi}_{int}(\beta) \nonumber \\
  &=&\frac{2J+1}{8\pi^2} \int d\Omega D^{J*}_{MK}(\Omega) R(\Omega)\Phi^{\pi}_{int}(\beta), \label{eq:prjwf}
\end{eqnarray} 
where, $P^{J}_{MK}$, $D^{J}_{MK}(\Omega)$ and ${R}(\Omega)$ denote the angular momentum projector,
the Wigner $D$ function and the rotation operator, respectively. The integrals over  three Euler
angles $\Omega$ are evaluated numerically. Then, we superpose the wave functions with different
quadrupole deformation $\beta$ and projection of angular momentum $K$ (GCM),  
\begin{eqnarray}
 \Psi^{J\pi}_{M\alpha} = \sum_{K=-J}^J\sum_{i=1}^N
 e_{Ki\alpha}\Phi^{J\pi}_{MK}(\beta_i),\label{eq:gcmwf} 
\end{eqnarray}
where, $N$ is a number of the basis wave functions to be superposed.
The coefficients $e_{Ki\alpha}$ and eigenenergy $E^{J\pi}_\alpha$ are obtained by solving the 
Hill-Wheeler equation~\cite{hill1953}, 
\begin{eqnarray}
 \sum_{K'i'}{H^{J\pi}_{KiK'i'}e_{K'i'\alpha}} &=& E^{J\pi}_\alpha
 \sum_{K'i'}{N^{J\pi}_{KiK'i'}e_{K'i'\alpha}},\\  
 H^{J\pi}_{KiK'i'} 
& =& \langle{\Phi^{J\pi}_{MK}(\beta_i)|H|\Phi^{J\pi}_{MK'}(\beta_{i'})}\rangle, \\
 N^{J\pi}_{KiK'i'} 
 &=& \langle{\Phi^{J\pi}_{MK}(\beta_i)|\Phi^{J\pi}_{MK'}(\beta_{i'})}\rangle.
\end{eqnarray}

\begin{figure}[h]
\centering\includegraphics[width=\linewidth,clip]{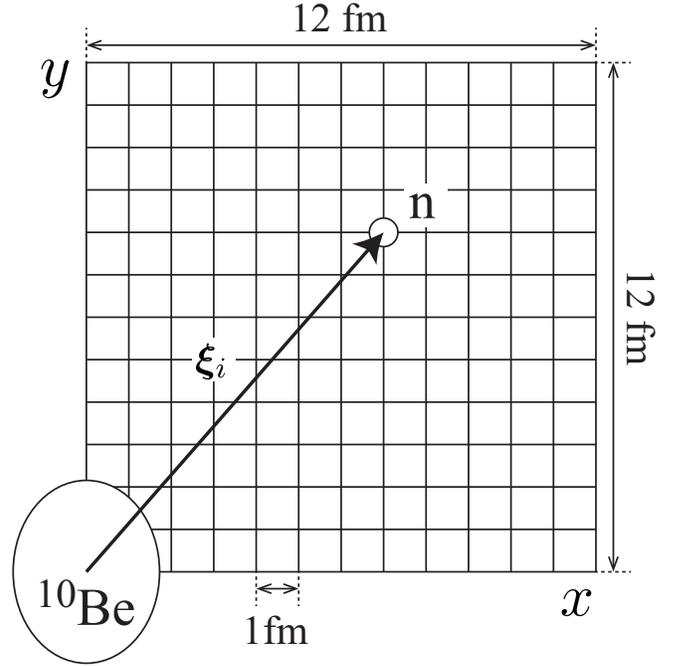}
\caption{A schematic illustration of the basis wave function for the $^{10}{\rm Be}+n$ system
 used in the AMD+RGM method.}  \label{fig:rgm}
\end{figure}

As explained later, the basis wave functions $\Phi^{J\pi}_{MK}(\beta_i)$ generated by the energy
variation are not sufficient to  describe the neutron halo of $^{11}{\rm Be}$. To incorporate with
the proper asymptotics of the halo wave function, we have introduced a set of wave functions as
additional basis. As schematically illustrated in Fig.~\ref{fig:rgm}, we have generated the
$^{10}{\rm Be}+n$ wave functions by placing the $^{10}{\rm Be}$ and $n$ on the grid points within
the 12 fm $\times $ 12 fm  size with 1 fm intervals. These wave functions may be represented as,
\begin{eqnarray}
 \Phi(\bm \xi_i,\chi_n) =\mathcal{A}
 \left\{\Phi_{{^{10}{\rm Be}}}(\bm -\frac{1}{11}\bm \xi_i)\varphi_{n}(\frac{10}{11}\bm
 \xi_i,\chi_n)\right\}. \label{eq:intrgm}
\end{eqnarray}
Here, the wave function of $^{10}{\rm Be}$ is the  intrinsic wave function
[Eq.~(\ref{eq:intwf})] obtained by the energy variation without constraint on the deformation parameter
$\beta$, and the valence neutron is described by a Gaussian wave packet [Eq.~(\ref{eq:singlewf})]
placed at ${10}/{11}\bm \xi_i$. Because $\Phi_{^{10}{\rm Be}}$ has approximate axial and reflection
symmetry, the relative coordinate $\bm \xi_i$ between $^{10}{\rm Be}$ and the valence neutron can be
restricted within the first quadrant of the $xy$-plane where the $y$-axis is the symmetry axis of
$\Phi_{^{11}{\rm Be}}$. Consequently, we have generated $13\times  
13\times 2= 338$ basis wave functions (number of grid points $\times$ neutron spin), which are
superposed after the angular momentum projection, 
\begin{align}
 \Psi^{J\pi}_{M\alpha} =& \sum_{K=-J}^J\biggl\{\sum_{i=1}^N
 e_{Ki\alpha}\Phi^{J\pi}_{MK}(\beta_i) \nonumber \\
 & +\sum_{i=1}^{169}\sum_{\chi_n=\uparrow,\downarrow}
 f_{Ki\chi_n\alpha}P^{J}_{MK}P^{\pi}\Phi(\bm \xi_i,\chi_n) \biggr\}. \label{eq:rgmwf}
\end{align}
The coefficients $e_{Ki\alpha}$, $f_{Ki\chi_n\alpha}$ and eigenenergy $E^{J\pi}_\alpha$ are
determined by solving the Hill-Wheeler equation again. We note that this method named AMD plus
resonating group method (AMD+RGM) has already been used to describe neutron halo of
$^{31}{\rm Ne}$~\cite{minomo2011, minomo2012, sumi2012}.

\subsubsection{Calculation of the valence neutron wave function}
We have extracted the valence neutron wave function (overlap amplitude) from
the microscopic wave functions of $^{10}{\rm Be}$ and $^{11}{\rm Be}$. For this purpose, firstly, we
calculate the overlap between the wave functions of $^{10}{\rm Be}$ and $^{11}{\rm Be}$. 
\begin{eqnarray}
 \psi(\bm r) = \sqrt{11}\left\{\Psi^{J^{\pi}}_{M}(^{10}{\rm Be})|
 \Psi^{1/2^+}_{M'}(^{11}{\rm Be})\right\}. \label{eq:ofunc1}
\end{eqnarray}
For simplicity, we assume that the wave functions of  $^{10,11}{\rm Be}$ are
described by the parity- and angular-momentum-projected wave functions given by
Eq.~(\ref{eq:prjwf}). Then, Eq. (\ref{eq:ofunc1}) reads,
\begin{eqnarray}
 \psi(\bm r)=\sum_{jl}C^{1/2M'}_{JM,jM'-M}u_{jl}(r)/r[Y_l(\hat r)\otimes \chi]_{jM'-M},\label{eq:ofunc2}
\end{eqnarray}
where, the overlap amplitude $u_{jl}(r)$ is defined as,
\begin{align}
 u_{jl}(r) =& \sum_k C^{1/2K'}_{JK'-k,jk}\sum_{p=1}^{11}(-)^pr\varphi^{(p)}_{jlk}(r)
  \nonumber\\
 &\times \frac{2J+1}{8\pi^2}\int d\Omega D^{J*}_{KK'-k}(\Omega) \det B^{(p)}(\Omega).\label{eq:ofunc3}
\end{align}
Here, $\varphi_{jlk}^{(p)}$ is the multi-pole decomposition of the single-particle wave packet
[Eq.~(\ref{eq:singlewf})],
\begin{eqnarray}
\varphi_p(\bm r) = \sum_{jlk}\varphi_{jlk}^{(p)}(r)[Y_l(\hat r)\otimes \chi]_{jk}.
\end{eqnarray}
$\det B^{(p)}(\Omega)$ is the determinant of the sub-matrix $B^{(p)}$ formed by removing $p$th
column from $B(\Omega)$. And $B(\Omega)$ is the ($10\times 11$)-dimension overlap matrix between
$^{10}{\rm Be}$ and $^{11}{\rm Be}$, that is defined as
$B(\Omega)_{ij}=<{\phi_i|R(\Omega)|\varphi_j}$ where $\phi_i$ and $\varphi_j$ are being the
single-particle wave packets of $^{10}{\rm Be}$ and $^{11}{\rm Be}$, respectively. Once the overlap
amplitude is calculated, its integral yields the spectroscopic factor, 
\begin{eqnarray}
 S_{jl} = \int_0^\infty dr\ |u_{jl}(r)|^2.
\end{eqnarray}
The derivation of the above formulae is explained in Ref.~\cite{kimura2017}. It is straightforward  
to extend them to the GCM and AMD+RGM wave functions given by Eqs.~(\ref{eq:gcmwf}) and
(\ref{eq:rgmwf}).

\section{Results and discussions}
\subsection{Static properties of ${}^{\rm 11}{\rm Be}$}\label{stat}
\begin{figure}[h]
\centering\includegraphics[width=\linewidth,clip]{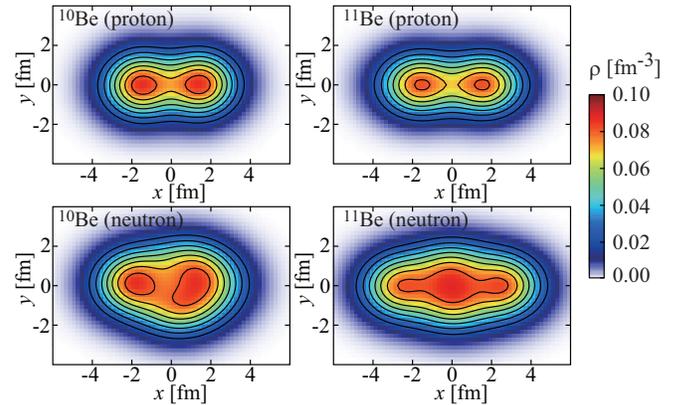}
 \caption{The proton and neutron density distributions calculated from the intrinsic wave functions
 which are the dominant component of the ground state. Upper (lower) panels show the proton
 (neutron) densities.}  \label{fig:intden}
\end{figure}

Fig.~\ref{fig:intden} shows the density profile of the  $^{10}{\rm Be}$ and $^{11}{\rm Be}$
intrinsic wave functions [Eq. (\ref{eq:intwf})] which are obtained by the energy variation and are
the dominant components of the ground state. 
\begin{figure}[h]
\centering\includegraphics[width=\linewidth,clip]{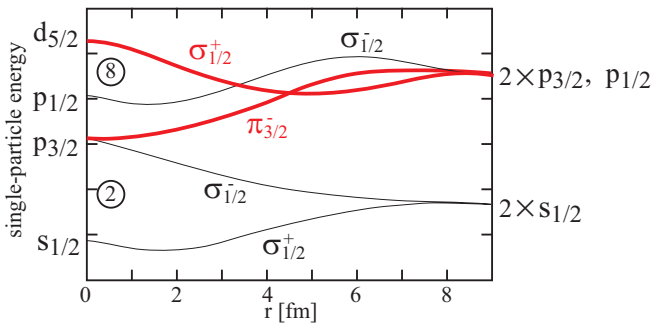}
 \caption{Schematic figure showing the behavior of the molecular orbits of $\alpha+\alpha$ cluster
 system as function of the inter-cluster distance. The orbits plotted with red lines show the
 molecular orbits ($\pi$- and $\sigma$-orbits) relevant to the low-lying states of $^{10}{\rm Be}$
 and $^{11}{\rm Be}$. This figure is reconstructed from Fig.~1 of Ref.~\cite{oertzen1997}.} 
\label{fig:mo}
\end{figure}
 As clearly seen, both nuclei have dumbbell-shaped proton density distributions which indicate the
 pronounced $\alpha+\alpha$ clustering. The valence neutrons (two valence neutrons of
 $^{10}{\rm Be}$ and three of $^{11}{\rm Be}$) occupy the so-called ``molecular orbits'' which are
 the single-particle orbits formed around $\alpha+\alpha$ cluster
 core~\cite{oertzen2006,oertzen1997,oertzen1997b,okabe1979}. In  both nuclei, two
 valence neutrons occupy the $\pi$-orbit, and the last valence neutron of $^{11}{\rm Be}$ occupies
 the $\sigma$-orbit. These molecular-orbit configurations are often denoted as $\pi^2$
 ($^{10}{\rm Be}$) and $\pi^2\sigma$ ($^{11}{\rm Be}$).  As discussed in
 Refs.~\cite{oertzen2006,oertzen1997,oertzen1997b,okabe1979,kanada1999,itagaki2000,itagaki2001,kanada2003x}, 
 the   $\pi$-orbit reduces the $\alpha+\alpha$  clustering, while the $\sigma$-orbit enhances it. This
 feature originates in the  single-particle energies of the molecular orbits as function of  the
 inter-cluster distance illustrated in Fig.~\ref{fig:mo}. The single-particle energy  of the
 $\pi$-orbit ($\sigma$-orbit) decreases (increases) as function of the inter-cluster  distance. As a
 result, the $\pi$-orbit ($\sigma$-orbit) favors weaker (stronger) $\alpha+\alpha$  clustering.
 Since, $^{11}{\rm Be}$ has an additional neutron in the $\sigma$-orbit, it manifests  more
 pronounced $\alpha+\alpha$ clustering than $^{10}{\rm Be}$  as seen in its proton density
 distribution  (Fig.~\ref{fig:intden}) and  larger quadrupole deformation parameter 
 (Table~\ref{tab:radii}). 
 These characteristics of $^{10}{\rm Be}$ and $^{11}{\rm Be}$ qualitatively agree with those discussed in
 the preceding studies~\cite{kanada1999,itagaki2000,itagaki2001,kanada2003x,neff2005}. We also note
 that the $\sigma$-orbit is a linear combination of the spherical  $sd$-shells, and hence, the
 valence neutron wave function (overlap amplitude) of $^{11}{\rm Be}$ should be, in general, an
 admixture of the $l=0$ and 2 components.   

\begin{figure}[h]
\centering\includegraphics[width=\linewidth,clip]{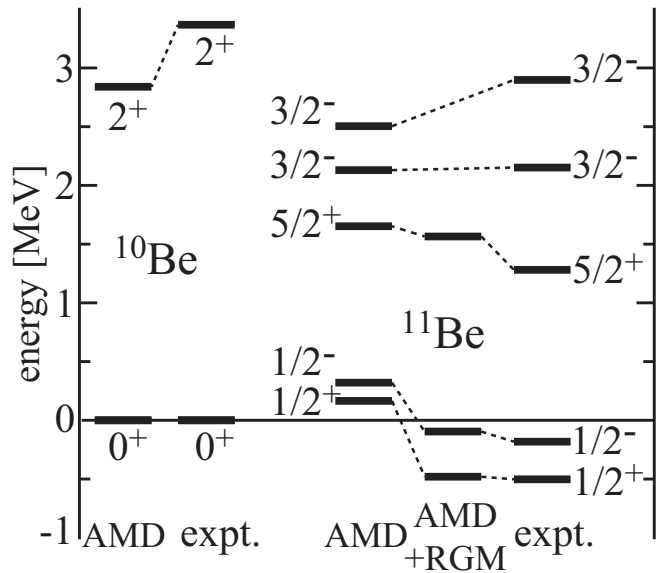}
 \caption{The calculated and observed excitation spectra of $^{10}{\rm Be}$ and $^{11}{\rm Be}$.
 The energy is relative to the ground state of $^{10}{\rm Be}$.} \label{fig:level}
\end{figure}

\begin{figure}[!h]
\centering
\includegraphics[width=\linewidth,clip]{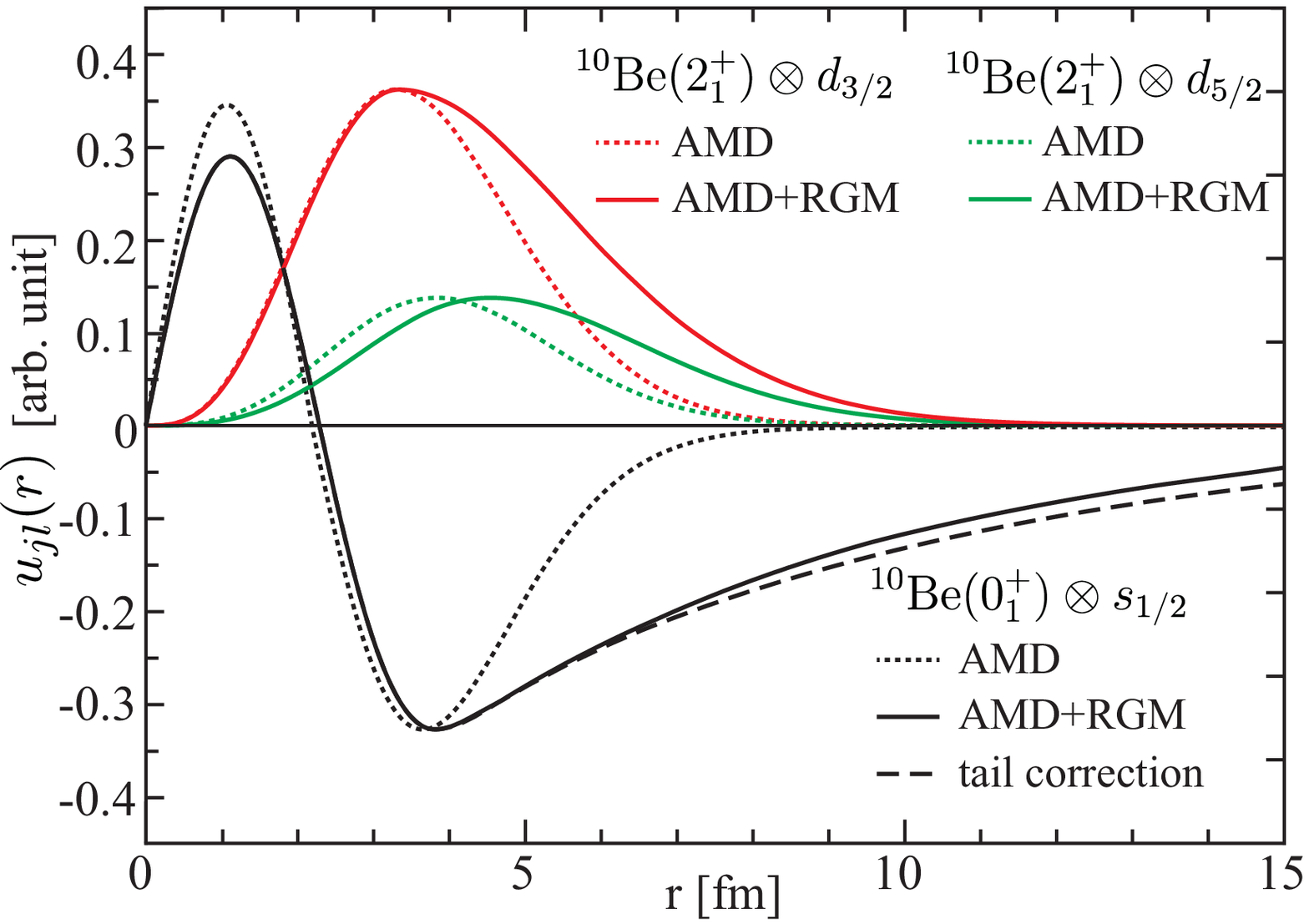}
 \caption{The overlap amplitudes of the valence neutron of $^{11}{\rm Be}$ calculated by the AMD and
 AMD+RGM. The amplitude is arbitrary scaled for the presentation.}\label{fig:ofunc}
\end{figure}

Fig.~\ref{fig:level} shows the excitation spectra of $^{10}{\rm Be}$ and $^{11}{\rm Be}$  obtained
by the GCM calculations (denoted as AMD in the figure). Contradictory to the observation, the 
adopted effective interaction does not bound $^{11}{\rm Be}$, although it gives the correct order of
the $1/2^\pm$ doublet and other excited states of $^{11}{\rm Be}$. The shortage of the binding energy is
due to the insufficient description of the asymptotics of the valence neutron wave function by the AMD
framework. 
 \begin{table*}[bht]
 \caption{The calculated and observed one-neutron separation energy ($S_n$), charge and point
  nucleon distribution radii ($\sqrt{\langle{r_{ch}^2}\rangle}$ and $\sqrt{\langle{r_{m}^2}\rangle}$). The second
  and third rows shows the results obtained by AMD and AMD+RGM calculations, respectively.   The
  calculated point neutron distribution radii ($\sqrt{\langle{r_{n}^2}\rangle}$) and the quadrupole
  deformation parameter for proton and neutron ($\beta_p$ and $\beta_n$) are also
  tabulated. The proton finite size effect is taken into account for the charge radii, while it is
  not for the neutron and matter radii.}\label{tab:radii}  
 \begin{center}
    \begin{tabular}{lcccccc|ccc}
     \hline
     &\multicolumn{6}{c}{AMD(+RGM)}&\multicolumn{3}{c}{expt.}\\\hline
     & $S_{n}$ & $\sqrt{\langle{r_{ch}^2}\rangle}$ & $\sqrt{\langle{r_m^2}\rangle}$ & $\sqrt{\langle{r_n^2}\rangle}$ 
     & $\beta_p$ & $\beta_n$  & $S_n$ & $\sqrt{\langle{r_{ch}^2}\rangle}$ & $\sqrt{\langle{r_m^2}\rangle}$\\
     $ $ &(MeV)&(fm)&(fm)&(fm)&&&(MeV)&(fm)&(fm)\\\hline
     $^{10}{\rm Be}$
     & 6.21 & 2.43 & 2.35 & $ 2.42$ & $0.60$ & $0.56$ 
     & 6.81 & $2.355(17)$ & $2.30(2)$\\\hline
     $^{11}{\rm Be}$
     & -0.22  & $2.54$ & $2.59$ & $2.71$ & $0.70$ & 0.69
     & 0.50 & $2.463(15)$ & 2.73(5)\\
     & 0.48  & $2.55$ & $2.71$ & 2.89 & $0.65$ & 0.62
     &       &      \\\hline     
    \end{tabular}
 \end{center}
 \end{table*}
The $s$-wave valence neutron wave function  shown in Fig.~\ref{fig:ofunc} (black dotted line) decays
at short distance and does not show halo nature, reflecting the fact that the AMD framework
approximates the valence neutron wave function by a single Gaussian. As a result, neither of the
binding nor large matter radius of $^{11}{\rm Be}$ are reproduced~\cite{Tanihata}.

To overcome this problem, we have performed the AMD+
RGM calculation which superposes the Gaussian
wave packets to describe the asymptotics of neutron halo. Fig.~\ref{fig:ofunc} shows
how the AMD+RGM drastically improves the results. The asymptotics of the $s$-wave 
($^{10}{\rm Be(0^+)}\otimes s_{1/2}$ channel) is greatly extended toward outside of the core
nucleus, and the neutron distribution radius is considerably increased compared to that calculated
by AMD (Table~\ref{tab:radii}).  On the contrary, the asymptotics of the $d$-wave
($^{10}{\rm Be(2^+)}\otimes d_{3/2}$ and $^{10}{\rm Be(2^+)}\otimes d_{5/2}$ channels) do not change
significantly. This may be due to  the centrifugal barrier in these channels which prevents the
long-ranged stretched asymptotics. Thanks to the improved asymptotics, the kinetic energy of the halo
orbit is reduced and the calculated one-neutron separation energy is now comparable with the
observed value.

An interesting side effect brought about by the AMD+\\
RGM is the reduction of the core deformation and
the decoupling between the core and valence neutron. As listed in Table~\ref{tab:radii}, the
quadrupole deformation of the proton distribution decreases in the AMD+RGM result ($\beta_p$=0.65)
compared to that of AMD ($\beta_p$=0.70). This implies that the coupling between the core
($^{10}{\rm Be}$) and the valence neutron is weakened, and the core polarization decreases. Indeed,
in the AMD+RGM result, the spectroscopic factor of the $s$-wave increases, while that of the $d$-wave
decreases compared to the AMD results.   
Thus, the AMD+RGM framework brings about a remarkable improvement of the neutron wave function and
offers a reasonable description of the neutron halo of $^{11}{\rm Be}$ with the deformed core
nucleus $^{10}{\rm Be}$. It is also noted that the overlap amplitudes obtained 
by the AMD+RGM (Fig.~\ref{fig:ofunc}) look consistent with those obtained by an ab-initio
calculations~\cite{NCSM,calci2016,bonaccorso2019}.

\newpage
\begin{table*}[bht]
 \caption{The calculated spectroscopic factors which are obtained from the integral of the overlap
 amplitudes shown in Fig.~\ref{fig:ofunc}. }\label{tab:sfac}   
 \begin{center}
    \begin{tabular}{cccc}
     \hline
     channel & AMD & AMD+RGM & expt.\\\hline
     $^{10}{\rm Be(0^+_1)}\otimes s_{1/2}$ & 0.46  & 0.82
	     & 1.0(2)~\cite{NakamuraPLB}, 0.77~\cite{Zwieglinski}, 0.74~\cite{Aumann}, 0.72(4)~\cite{Fukuda} \\\hline     
     $^{10}{\rm Be(2^+_1)}\otimes d_{3/2}$ & 0.38  & 0.18 & 0.18~\cite{Aumann} \\\hline     
     $^{10}{\rm Be(2^+_1)}\otimes d_{5/2}$ & 0.11  & 0.06 &  \\\hline     
    \end{tabular}
 \end{center}
\end{table*}
For later use, we further improved the asymptotics of the valence neutron wave function to be
consistent with the observed one-neutron separation energy, $S_n=0.50$ MeV. The $s$-wave overlap
function $u_0(r)$ calculated from AMD+RGM is smoothly connected to the exact asymptotics of
$A\exp(-\kappa r)$ where  $\kappa=\sqrt{2\mu S_n}/\hbar$ and $\mu$ is being the reduced mass for
$^{10}{\rm Be}+n$ system. The amplitude $A$ and the matching distance $a$ are determined from the
condition, 
\begin{align}
 \left.\frac{d}{dr}u_0(r)\right|_{r=a} =  \left.\frac{d}{dr}A\exp(-\kappa r)\right|_{r=a}. 
\end{align}
The overlap function with the tail correction, thus obtained, is shown by the dashed line in
Fig.~\ref{fig:ofunc}. This overlap function is used as an input to calculate various reaction 
observables in the breakup of $^{11}$Be on a heavy target in the subsequent sections.

\subsubsection{Wave function inputs for reaction calculations}
The bound state, single-particle radial wave function of $^{11}$Be has been built from a $^{10}{\rm
Be(0^+)} \otimes$ 1$s_{1/2}\nu$ configuration with a one neutron separation energy ($S_{n}$) of 0.50
MeV. It is also known that the contribution from the $d$-wave configuration is an order of magnitude
lower than the $s$-wave and so any admixture (along with a low spectroscopic factor, \textit{cf.} Table~\ref{tab:sfac}) would not be
perceptible in reaction observables (see e.g. Refs. \cite{RCNPA, Nunes, NakamuraPLB}). 

\begin{figure}[h]
	\centering\includegraphics[width=\linewidth,clip]{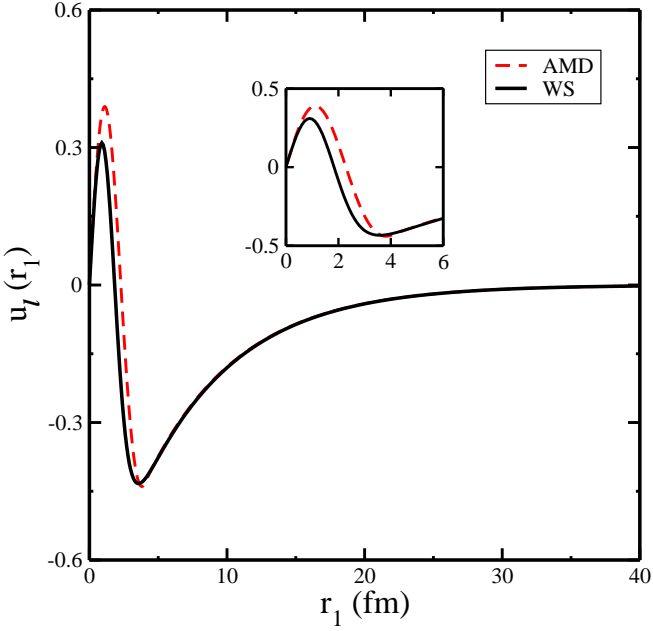}
	\caption{The normalized bound state radial wave functions of $^{11}$Be in WS (solid line) and AMD (dashed line) models. The inset shows the wave functions in the nuclear interior.}
	\label{wfn}
\end{figure}

The bound state radial wave function of $^{11}$Be was constructed in two ways. The first by
considering a Woods-Saxon potential of 70.99 MeV, radius and diffuseness parameters as 1.15 fm and
0.50 fm, respectively, so as to reproduce the one-neutron separation energy of 0.50 MeV. The other
is by using the overlap wave function obtained by the AMD + RGM framework with tail correction, as described earlier. In all subsequent sections we refer to this wave function as the AMD wave function itself. 
The two wave functions are compared in Fig.~\ref{wfn}. The solid and dashed lines show the
phenomenological WS and the microscopic AMD wave functions, respectively.  
%Wherever, comparison has been made with available data, our calculations have been multiplied with
%spectroscopic factor of 1. %\cite{Auton}. 

\subsubsection{Matter and charge radii in the cluster model}

It will be interesting to calculate the matter and charge radii in the cluster model \cite{rm}, with the phenomenological WS and AMD wave functions as inputs and compare them with those obtained in the previous section (Table~\ref{tab:radii}). 
The mean square matter radius $\langle r^2 \rangle_{A_a}$ and charge radius $\langle r^2_{ch} \rangle_{A_a}$ of a dicluster nucleus of mass number $A_a$ and charge $Z_a$ (consisting of subclusters $A_b, Z_b$ and  $A_c, Z_c$) can be written as \cite{rm} 

\begin{eqnarray}
\langle r^2 \rangle_{A_a} = \dfrac{A_b}{A_a} \langle r^2 \rangle_{A_b} + \dfrac{A_c}{A_a} \langle r^2 \rangle_{A_c} + \dfrac{A_b * A_c}{A_a^2} \langle R^2 \rangle ,
\label{eq:matter1}
\end{eqnarray}
and 
\begin{eqnarray}
\langle r^2_{ch} \rangle_{A_a} = \dfrac{Z_b}{Z_a} \langle r^2 \rangle_{A_b} + \dfrac{Z_c}{Z_a} \langle r^2
\rangle_{A_c} &+& \dfrac{\langle R^2 \rangle}{Z_a}\bigg(Z_b . \bigg(\dfrac{{A_c}}{{A_a}}\bigg)^2 
\nonumber\\
&+& Z_c . \bigg(\dfrac{{A_b}}{{A_a}}\bigg)^2\bigg),
\label{eq:charge1}
\end{eqnarray}
\noindent
respectively. In Eqs. (\ref{eq:matter1}) and (\ref{eq:charge1}), $\langle R^2 \rangle = \langle u_l(r_1) \vert r_1^2 \vert u_l(r_1) \rangle.$

In our case, \textit{b} and \textit{c} are the core ($^{10}$Be) and the valence neutron of the projectile respectively. If we neglect the second term of Eq. (\ref{eq:matter1}), then the mean square matter radius can be written as
\begin{eqnarray}
\langle r^2 \rangle_{A_a} = \dfrac{A_b}{A_a} \langle r^2 \rangle_{A_b} + \dfrac{A_b * A_c}{A_a^2} \langle R^2 \rangle.
\label{eq:matter}
\end{eqnarray} 
Furthermore, given that $Z_c = 0$, the mean square charge radius simplifies to 
\begin{eqnarray}
\langle r^2_{ch} \rangle_{A_a} = \dfrac{Z_b}{Z_a} \langle r^2 \rangle_{A_b} + \dfrac{\langle R^2 \rangle}{Z_a}\bigg(Z_b . \bigg(\dfrac{A_c}{A_a}\bigg)^2\bigg) .
\label{eq:charge}
\end{eqnarray}

\begin{table}[h]
 \caption{Root mean square matter ($\sqrt{\langle r^2 \rangle_{A_a}}$) and charge ($\sqrt{\langle r^2_{ch} \rangle_{A_a}}$) radii of $^{11}$Be in the cluster model.}\label{tab:deformation} 
 \begin{center}
%  \begin{ruledtabular}
   \begin{tabular}{lccccc}
    \hline 
       &WS (fm)& AMD (fm)& expt. (fm) & 
 \\\hline
$\sqrt{\langle r^2 \rangle_{A_a}}$ &   2.79  & 2.78  & 2.73(0.05)\cite{Tanihata}   \\

\hline

$\sqrt{\langle r^2_{ch} \rangle_{A_a}}$&   2.35 & 2.35  & 2.463(15)\cite{Zakova}  \\
\hline
% & & & 2.463(0.015)\cite{Angeli} \\
% &&&2.466(15)\cite{Krieger} \\
   \end{tabular}
%  \end{ruledtabular}
 \end{center}
\end{table}

For our calculations, we have used the size of the charged core, $\sqrt{\langle r^2 \rangle_{A_b}}$ = 2.28 fm. \cite{AlkhaliliPRL}. 
Evidently, the WS and the AMD wave functions directly enter the calculation of the radii only through $\langle R^2 \rangle$ term in Eqs. (\ref{eq:matter} - \ref{eq:charge}). Thus any difference between the phenomenological WS wave function and the microscopic AMD will be reflected in $\langle R^2 \rangle$.  We have also used a spectroscopic factor of 0.82 (in Table~\ref{tab:sfac}), obtained in the structural calculations shown earlier.

In Table 3, we show that our calculated root mean square matter and charge radii using both the WS and AMD wave functions compare well with the available data.  The matter radius calculated from other models such as relativistic mean field model (2.52 fm \cite{Wang}), Glauber model (2.76(0.03) fm \cite{Suhelahmed}),
% 2.73(0.05) fm \cite{Ozawa2} 
and fermionic molecular dynamics model (2.80 fm \cite{Krieger}) also agree well with our results. Similarly, the charge radius deduced from the fermionic molecular dynamics (2.38 fm \cite{Krieger}) and no core shell model (2.37(11) fm \cite{NCSM}) agree well with the WS and AMD results.

\subsection{Reaction observables}
\label{Reaction.sec}

\subsubsection{Triple differential cross section}
\begin{figure}[h]
\centering\includegraphics[width=\linewidth,clip]{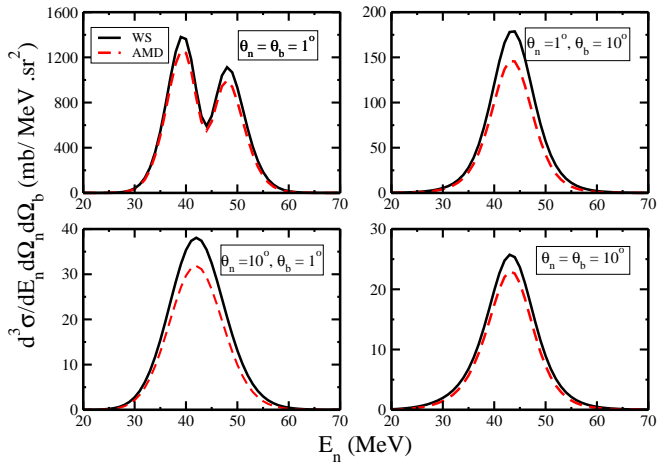}
\caption{Triple differential cross section for the breakup of $^{11}$Be on $^{197}$Au at a beam energy of 44 MeV/nucleon. The solid and dashed lines correspond to WS and AMD calculations, respectively. For more details see text.}
\label{triple}
\end{figure}

In Fig.~\ref{triple} we plot the triple differential cross sections in the breakup of $^{11}$Be on $^{197}$Au target at a beam energy of 44 MeV/nucleon as a function of the neutron energy (E$_n$), for four different combinations of the neutron angle ($\theta_n$) and the angle of the charged fragment ($\theta_b$). 
Given that the spectroscopic factor of the s-wave is close to unity \cite{ Aumann, NakamuraPLB, Zwieglinski, Fukuda}, our subsequent breakup calculations also take it as unity.

As expected the cross sections are indeed larger at small scattering angles as the breakup is Coulomb dominated.  
We observe that the calculations obtained from both the WS and AMD wave functions are similar. 
 Given that triple differential cross sections are exclusive reaction observables, this similarity in results builds up an expectation that other reaction observables may also not be too different while using these wave functions. This is because other inclusive reaction observables can be obtained from the triple differential cross section after performing suitable integrations over various kinematic parameters.
 
\subsubsection{Neutron energy distribution}
\begin{figure}[h]
\centering\includegraphics[width=\linewidth,clip]{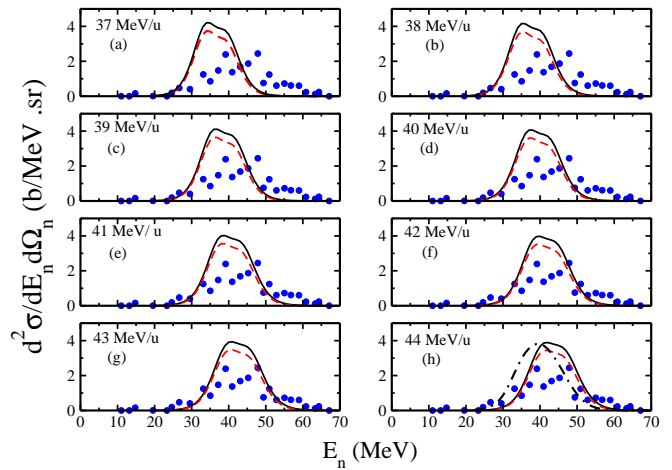}
	\caption{Neutron energy distributions in the Coulomb breakup of $^{11}$Be on a gold target at beam energies ranging from 37 to 44 MeV/nucleon, calculated using WS (solid line) and AMD (dashed line) wave functions. The experimental data are shown by solid circles are from Ref. \cite{Annened}. In figure (h), the dot-dashed line shows the average of all the WS results performed at different beam energies.}
\label{ned_comparison}
\end{figure}

We now calculate the neutron energy distribution in the breakup of $^{11}$Be on $^{197}$Au and compare it with existing experimental data (solid circles) \cite{Annened}, at $\theta_n = 1^\circ$. Incidentally, the beam energy for the same experimental data was not unique and was in the range of 36.9 to 44.1 MeV/u. To take care of this variation in our calculation, in Fig.~\ref{ned_comparison} we have shown the neutron energy distribution using the WS (solid line) and AMD (dashed line) wave functions, calculated at a series of beam energies ranging from 37 - 44 MeV/u. In the same figure [Fig.~\ref{ned_comparison}(h), dot-dashed line], we have also plotted the average of all the WS results performed at different beam energies. 

With the progress in current radioactive ion beam facilities, it would indeed be interesting to perform these exclusive measurements as that could constrain the spread in the data seen in Fig.~\ref{ned_comparison}.

\subsubsection{Parallel momentum distribution}
We now turn our attention to the PMD of the charged fragment in the breakup of $^{11}$Be on $^{181}$Ta at a beam energy of 63 MeV/u. The width of this distribution is also a measure of the size of the nucleus in coordinate space. From a statistical model calculation \cite{Goldhaber}, the width can be given by $\Delta^2 = \Delta_0^2  \big[A_b(A_a - A_b)/A_a $), where $A_a$ and $A_b$ are the mass numbers of the projectile and fragment, respectively and $\Delta_0$ ($\approx$ 80 MeV/c) has a constant value and it is also known to be independent of target mass and beam energy. This approximation suggests that the width of the $^{10}$Be distribution in the breakup of $^{11}$Be could be approximately 80 MeV/c. However, the experimental FWHM of the PMD for the \textit{s}-wave configuration has been found to be $43.6 \pm1.1 $ MeV/c \cite{KellyPMD}. 

\begin{figure}[h]
\centering\includegraphics[width=\linewidth,clip]{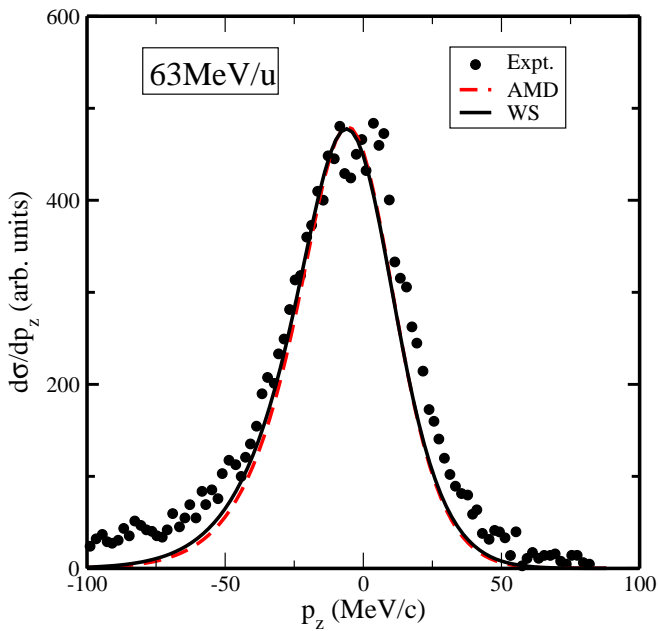}
	\caption{The parallel momentum ($p_z$) distribution of $^{10}$Be in the elastic Coulomb breakup of $^{11}$Be on $^{181}$Ta at $63$ MeV/u in the rest frame of the projectile. The solid and dashed lines correspond to WS and AMD calculations, respectively and the experimental data, shown by solid circles, are from Ref. \cite{KellyPMD}.}
\label{pmd}
\end{figure}

In Fig.~\ref{pmd} we show our fully quantum mechanical calculation of the PMD of the charged core ($^{10}$Be) fragments emitted during the elastic Coulomb breakup of $^{11}$Be on $^{181}$Ta at a beam energy of 63 MeV/u, in the rest frame of the projectile. The solid and the dashed lines correspond to our calculations with WS and AMD wave functions, respectively and are normalized to the peak of the data. The FWHM for both the theoretical (WS and AMD) calculations are 42 MeV/c which is in good agreement with the experimental value of $43.6 \pm1.1 $ MeV/c from Ref.\cite{KellyPMD}.

\subsubsection{Relative energy spectrum and dipole response}

\begin{figure}[h]
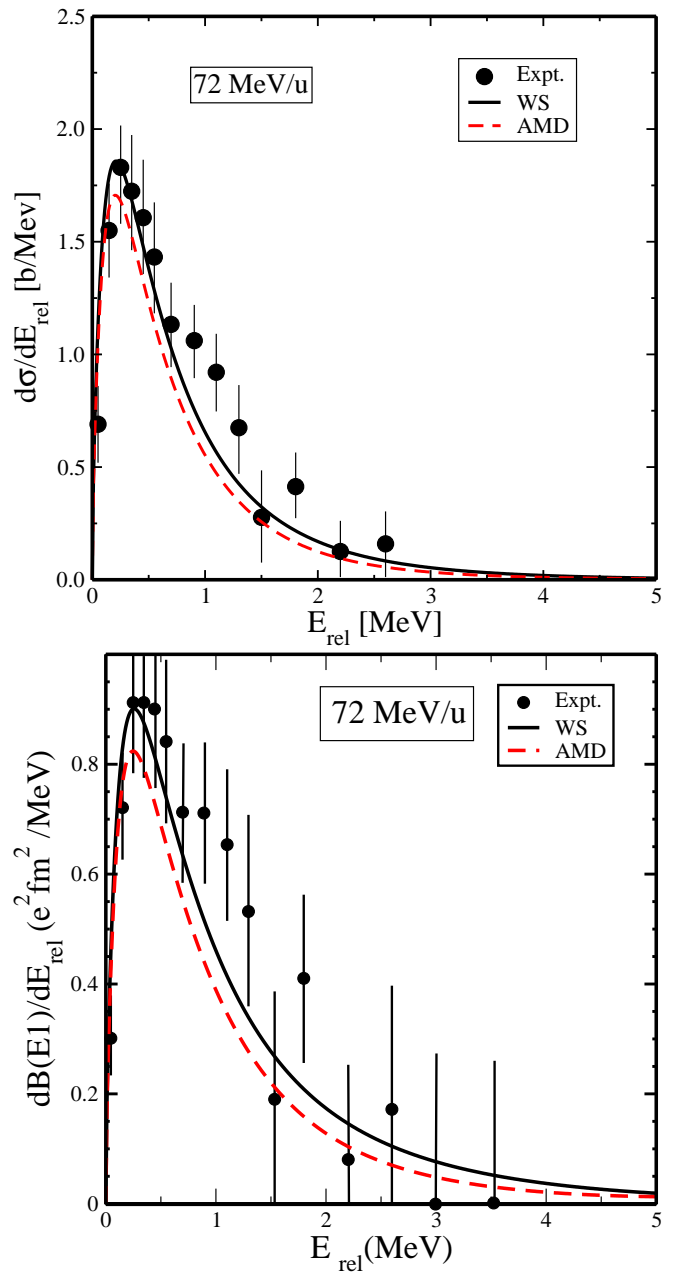

\centering\includegraphics[width=\linewidth,clip]{rel.eps}
\centering\includegraphics[width=\linewidth,clip]{dbde_72.eps}
	\caption{The relative energy spectra in the elastic Coulomb breakup of $^{11}$Be on $^{208}$Pb at 72 MeV/u (upper panel) and dipole strength distribution (lower panel). Calculations with the WS and the AMD wave functions are shown by solid and dashed lines, respectively. The solid circles show the experimental data from Ref. \cite{NakamuraPLB}.}
\label{rel}
\end{figure}

We now continue our efforts in testing the WS and the AMD wave functions by calculating the relative energy spectrum and the dipole strength distribution of $^{11}$Be on a heavy target. In Fig.~\ref{rel}, we have shown the relative energy spectrum in the elastic Coulomb breakup of $^{11}$Be on a $^{208}$Pb target at beam energy of 72 MeV/u (upper panel) and dipole strength distribution (lower panel) and compare our results with the available experimental data \cite{NakamuraPLB}. 
The solid and the dashed lines represent FRDWBA calculations using the WS and the AMD wave functions, respectively, while the experimental data are shown by solid circles. We see that both the WS and AMD results are able to reproduce the peak positions of the relative energy spectra and the dipole distribution quite well. Contributions at higher relative energies are dominated by the nuclear breakup \cite{RCPPNP}.

 The total one neutron removal cross sections ($\sigma_{n}$), obtained by integrating the relative energy spectrum with the WS and AMD wave functions, are 1.76 b and 1.51 b, respectively and the corresponding experimental value is 1.8 $\pm$ 0.4 b \cite{NakamuraPLB}. 

%In Ref. \cite{Ogata}, the reported one neutron removal cross section ($\sigma_{-n}$) by a Pb target at a beam energy of 250 MeV/u is 1.12 b. They have used the continumm-dicretized coupled-channel method with eikonal theory for the elastic breakup cross section ($\sigma_{eb}$) in combination with the eikonal theory for the one neutron stripping cross section ($\sigma_{str}$) to calculate the one neutron removal cross section ($\sigma_{-n}$ = $\sigma_{eb}$ + $\sigma_{str}$) \cite{Ogata,OgataCDCC1}.  The discrepancy between the experimental and the theoretical measurements can be due to the nuclear breakup effects. 

Again the total $B(E1)$ value for $^{11}$Be determined experimentally at a beam energy of 72 MeV/u is 1.3 $\pm$ 0.3 $e^2fm^2$ \cite{NakamuraPLB} and the theoretical values obtained by integrating the lower panel of Fig.~\ref{rel} for WS and AMD calculations are 1.17 and 0.97 $e^2fm^2$, respectively. We emphasize that our post form FRDWBA considers the breakup process to be a one-step process \cite{RCPPNP} and hence the agreement of our calculations with the experimental data is a direct proof that the enhanced dipole strength at low energies is due to the breakup of nucleus into the continuum and not because of any soft dipole resonance. A similar conclusion was also reached by the authors of \cite{Fukuda}. 

Interestingly, in an extreme single-particle model, provided other excited bound states do not contribute to the dipole transition, the total $B(E1)$ is known to be proportional to the mean square radius of the valence nucleon ($\langle r^2_{\nu} \rangle$) \cite{BertulaniNPA, BertulaniPRC,Esben,Nagarjan,Typel2005}, via $B(E1)$ = (3/4$\pi$)$(Z_a \textit{e}/A_a)^2$ $\langle r^2_{\nu} \rangle$. Under this approximation, using the theoretical estimates of $B(E1)$ obtained earlier, the root mean square radius of the valence neutron ($\sqrt{\langle r^2_{\nu} \rangle}$) with the WS and the AMD wave functions turns out to be 6.08 fm and 5.54 fm, respectively. These do compare well with experimental estimate of 6.4 $\pm$ 0.7 fm in Ref. \cite{NakamuraPLB}.

\section{Conclusions}
We have investigated the static properties and reaction observables of $^{11}$Be breaking in the
presence of Coulomb field of $^{208}$Pb. For our theoretical calculations we have combined the AMD
framework  and the fully quantum mechanical FRDWBA model.

We have used the AMD to describe the structure and the valence neutron wave function of
$^{11}{\rm Be}$. To incorporate the long-ranged asymptotics of the halo wave function, an extended
AMD framework named AMD+RGM has also been adopted. The AMD+RGM has drastically improved the
asymptotics of halo wave function, and plausibly described various properties such as 
charge and neutron distribution radii, one neutron separation energy and excitation spectrum. 

The valence neutron wave function calculated by AMD+RGM was used as an input of the FRDWBA model  to
describe the dynamical properties of $^{11}{\rm Be}$. The advantage of the method over other
first-order perturbative theories is that it requires only the ground state wave function of the
projectile as an input and includes all orders of the electromagnetic interaction between the
fragments and the target.

Apart from the many body AMD, a phenomenological Woods-Saxon wave functions has also been used for
the purpose of comparison. We have shown that the static properties calculated with these wave
functions, mainly the matter and charge radii of $^{11}$Be, are in good agreement with the available
data. This gave us the confidence to calculate several reaction observables in the breakup of
$^{11}$Be on heavy targets. Several observables like the triple differential cross section, neutron
energy distribution, parallel momentum distribution, relative energy spectrum and dipole strength
distribution are presented and compared with experimental data, wherever available. The upshot of
this method is that the same input is used to calculate various exclusive and inclusive observables.  
Given the validity of our method in the low mass region, it would also be interesting to extend our
calculations to the deformed medium mass region of the nuclear chart where experimental data are
scarce.

%The most interesting finding of this work is the total $B(E1)$ value and thereafter, the root mean square radius of the valence nucleon from the total $B(E1)$ value. The root mean square radius of the valence neutron along with charge and matter radius results strongly validate the idea of s-wave halo structure of $^{11}$Be. 

\section*{Acknowledgment}
This work was supported by the Scheme for Promotion of Academic and Research Collaboration (SPARC/2018-2019/P309/SL), Ministry of Education, India. M.K. acknowledges the support from JSPS KAKENHI Grant No. 19K03859, the collaborative research programs 2020, Information Initiative Center at Hokkaido University and  the COREnet program at the RCNP, Osaka University. M.D. acknowledges a doctoral research  fellowship from the Ministry of Education, India.


%\bibliographystyle{ptephy}
%\bibliography{sample}

\begin{thebibliography}{9}
 \bibitem{Tanihata halo} I. Tanihata, \textit{et al.}, {Phys. Rev. Lett.} \textbf{55}, 2676 (1985).
	 
 \bibitem{Shubh} Shubhchintak, R. Chatterjee, {Phys. Rev. C} \textbf{90}, 017602 (2014).

 \bibitem{Aumann} T. Aumann, T. Nakamura, {Phys. Scr.}, \textbf{T152}, 014012 (2013).

 \bibitem{Tanihata} I. Tanihata, \textit{et al.}, {Phys. Lett. B} \textbf{206}, 592 (1988).

 \bibitem{Charge radii} W. N\"{o}rtersh\"{a}user, \textit{et al.}, {Phys. Rev. Lett.} \textbf{102}, 062503 (2009).

 \bibitem{Wang}J. S. Wang, \textit{et al.}, {Nucl. Phys. A} \textbf{691}, 618 (2001).

 \bibitem{Suhelahmed} Suhel Ahmed, A. A. Usmani, Z. A. Khan, {Phys. Rev. C} \textbf{96}, 064602 (2017).

%\bibitem{Ozawa2}A. Ozawa, \textit{et al.}, {Nucl. Phys. A} \textbf{693}, 32 (2001).

 \bibitem{Krieger} A. Krieger, \textit{et al.}, {Phys. Rev. Lett.} \textbf{108}, 142501 (2012).

 \bibitem{NCSM} C. Forss\'{e}n, E. Caurier, P. Navr\'atil, {Phys. Rev. C} \textbf{79}, 021303 (2009).

 \bibitem{Warner} R. E. Warner, \textit{et al.}, {Phys. Rev. C} \textbf{64}, 044611 (2001).

 \bibitem{Capel2018} P. Capel, \textit{et al.}, {J. Phys.: Conf. Ser.} \textbf{1023}, 012010 (2018).

 \bibitem{Lei} A. Leistenschneider, \textit{et al.}, {Phys. Rev. Lett.} \textbf{86}, 5442 (2001).

 \bibitem{softdipole1} J. Chambers, \textit{et al.}, {Phys. Rev. C} \textbf{50},  R2671 (1994).

 \bibitem{softdipole2} Y. Suzuki, K. Ikeda, H. Salto, {Prog. Theor. Phys.} \textbf{83}, 180 (1990).
 
 \bibitem{NakamuraPLB} T. Nakamura, \textit{et al.}, {Phys. Lett. B} \textbf{331}, 296 (1994).

 \bibitem{NakamuraPRL} T. Nakamura, \textit{et al.}, {Phys. Rev. Lett.} \textbf{83}, 1112 (1999).

 \bibitem{6He}S. Nakayama, \textit{et al.}, {Phys. Rev. Lett.} \textbf{85}, 262 (2000).

 \bibitem{11li} R. Kanungo, I. Tanihata, C. Samanta, {Prog. Theor. Phys.} \textbf{102}, 1133 (1999).

 \bibitem{BertulaniNPA} C.A. Bertulani, G. Baur, M.S. Hussein, {Nucl. Phys. A} \textbf{526}, 751 (1991).


 \bibitem{BertulaniPRC}C.A. Bertulani, A. Sustich, {Phys. Rev. C} \textbf{46}, 6 (1992).

 \bibitem{Goriely} S. Goriely, {Phys. Lett. B} \textbf{436}, 10 (1998).

 \bibitem{BertulaniEPJA} C.A. Bertulani, {Eur. Phys. J. A} \textbf{55}, 240 (2019).
 
 \bibitem{kanada2003} Y. Kanada-En'yo, M. Kimura, H. Horiuchi, Comptes Rendus Physique
	 {\bf 4}, 497 (2003).
 \bibitem{kanada2012} Y. Kanada-En'yo, M. Kimura, A. Ono, Prog. Theor. Exp. Phys. 
	 {\bf 2012}, 01A202 (2012).
 \bibitem{kimura2016} M. Kimura, T. Suhara, Y. Kanada-En'yo, Eur. Phys. J. A {\bf 52}, 373 (2016).
 \bibitem{Fuchs} H. Fuchs, {Nucl. Instrum. Methods} \textbf{200}, 361 (1982).
 \bibitem{RCPPNP} R. Chatterjee, R. Shyam, Prog. Part. Nucl. Phys. \textbf{103}, 67 (2018).
 \bibitem{ManjuEPJ} Manju, J. Singh, Shubhchintak, R. Chatterjee, {Eur. Phys. J. A} \textbf{55}, 5 (2019).

 \bibitem{berger1991} J.F. Berger, M. Girod,  D. Gogny, Computer Physics Communications {\bf 63},
	 365 (1991).
 \bibitem{homma2015} H. Homma, M. Isaka, M. Kimura, Phys. Rev. {\bf 91}, 014314 (2015).
 \bibitem{kimura2004} M. Kimura, Phys. Rev. C {\bf 69}, 044319 (2004).
 \bibitem{kimura2012} M. Kimura, R. Yoshida, M. Isaka, Prog. Theor. Phys. {\bf 127}, 287 (2012).
 \bibitem{hill1953} D.L. Hill,  J.A. Wheeler, Phys. Rev. {\bf 89} 1102 (1953).
 \bibitem{minomo2011} K. Minomo, T. Sumi, M. Kimura, K. Ogata, Y.R. Shimizu, M. Yahiro,
	 Phys. Rev. C {\bf 84}, 034602 (2011).
 \bibitem{minomo2012} K. Minomo, T. Sumi, M. Kimura, K. Ogata, Y.R. Shimizu, M. Yahiro,
	 Phys. Rev. Lett. {\bf 108} 052503 (2012).
 \bibitem{sumi2012} T. Sumi, K. Minomo, S. Tagami, M. Kimura, T. Matsumoto, K. Ogata, Y.R. Shimizu, M. Yahiro, Phys. Rev. C {\bf 85}, 064613 (2012).
 \bibitem{kimura2017} M. Kimura, Phys. Rev. C {\bf 95}, 034331 (2017).
 \bibitem{oertzen2006} W. von Oertzen, M. Freer, Y. Kanada-En'yo, Phys. Rep. {\bf 432}, 43 (2006).
 \bibitem{oertzen1997} W. von Oertzen, II Nuovo Cimento A {\bf 110}, 895 (1997)
 \bibitem{oertzen1997b} W. von Oertzen, Z. Physik A {\bf 357}, 355 (1997).
 \bibitem{okabe1979} S. Okabe, Y. Abe, H. Tanaka, Prog. Theor. Phys. {\bf 57}, 866 (1979).
 \bibitem{kanada1999} Y. Kanada-En’yo, H. Horiuchi, A. Dote, Phys. Rev. C {\bf 64}, 0564304 (1999).
 \bibitem{itagaki2000} N. Itagaki, S. Okabe, Phys. Rev. C {\bf 61}, 044306 (2000).
 \bibitem{itagaki2001} N. Itagaki, S. Okabe, K. Ikeda, I. Tanihata, Phys. Rev. C {\bf 64} 014301 (2001).
 \bibitem{kanada2003x} Y. Kanada-En’yo, H. Horiuchi, Phys. Rev. C {\bf 68}, 014319 (2003).
 \bibitem{neff2005} T. Neff, H. Feldmeier, R. Roth, Nucl. Phys. A {\bf 752}, 321 (2005).

 \bibitem{calci2016} A. Calci, P. Navr\'{a}til, R. Roth, J. Dohet-Eraly, Sofia Quaglioni,
	 Guillaume Hupin, Phys. Rev. Lett. {\bf 117}, 242501 (2016).
 \bibitem{bonaccorso2019} A. Bonaccorso, F. Cappuzzello, D. Carbone, M. Cavallaro, G. Hupin, P. Navr\'{a}til, S. Quaglioni, Phys. Rev. C {\bf 100}, 024617 (2019).
		  
\bibitem{RCNPA} R. Chatterjee, P. Banerjee, R. Shyam, {Nucl. Phys. A} \textbf{675}, 477 (2000).

\bibitem{Nunes} F. M. Nunes, I. J. Thompson, R. C. Johnson, {Nucl. Phys. A} \textbf{596}, 171 (1996).
		  
 \bibitem{Zwieglinski} B. Zwieglinski, W. Benenson, R.G.H. Robertson, W.R.Coker,
	 Nucl. Phys. A {\bf 315}, 124 (1979). 
	 
\bibitem{Fukuda} N. Fukuda, \textit{et al.}, {Phys. Rev. C} \textbf{70}, 054606 (2004).


\bibitem{rm} A. Mason, R. Chatterjee, L. Fortunato, A. Vitturi, {Eur. Phys. J. A} \textbf{39}, 107 (2009).

\bibitem{Zakova} M. \v{Z}\'{a}kov\'{a}, \textit{et al.}, {J. Phys. G.} \textbf{37}, 055107 (2010).

\bibitem{AlkhaliliPRL} J. S. Al-Khalili, J. A. Tostevin, {Phys. Rev. Lett.} \textbf{76}, 3903 (1996).

	 
%\bibitem{Auton} D.L. Auton, {Nucl. Phys. A} \textbf{157}, 305 (1970).



\bibitem{Annened} R. Anne, \textit{et al.}, {Nucl. Phys. A} \textbf{575}, 125 (1994).

\bibitem{Goldhaber} A. S. Goldhaber {Phys. Lett. B} \textbf{53}, 306 (1974).

\bibitem{KellyPMD} J. H. Kelly, \textit{et al.}, {Phys. Rev. Lett.} \textbf{74}, 30 (1995).

%\bibitem{Ogata} K. Yoshida,  T. Fukui, K. Minomo, K. Ogata, {Prog. Theor. Exp. Phys.} \textbf{2014}, 053D03 (2014).

%\bibitem{OgataCDCC1} K. Ogata, \textit{et al.}, {Phys. Rev. C} \textbf{68}, 064609 (2003).


\bibitem{Esben}J. H. Esbensen, G.F. Bertsch, {Nucl. Phys. A} \textbf{542}, 310 (1992).
 
\bibitem{Nagarjan} M.A. Nagarajan, S. M. Lenzi, A. Vitturi, {Eur. Phys. J. A} \textbf{24}, 63 (2005).

\bibitem{Typel2005} S. Typel, G. Baur, {Nucl. Phys. A} \textbf{759}, 247 (2005).



\end{thebibliography}
%
% once the .bbl file has been generated then place the text in your article.

\vspace{0.2cm}
\noindent

%This is added by T. Yoneya (editor-in-chief) on 2020/07/09.
%\balance

\let\doi\relax

\appendix

\end{document}